\documentclass[useAMS,usenatbib]{mn2e} 

\usepackage{times}
\usepackage{graphicx}
\usepackage{amssymb}
\usepackage{color}
\usepackage[normalem]{ulem}

\newcommand{\ion}[2]{{#1~\small#2}}

\newcommand{\ha}{H$\alpha$}
\newcommand{\hb}{H$\beta$}

\newcommand{\kms}{$\rm km~s^{-1}$}

\newcommand{\eso}{ESO137$-$001}

\title[A MUSE study of \eso]{MUSE sneaks a peek at extreme ram-pressure stripping events. II. 
The physical properties of the gas tail of ESO137-001.}

\author[Fossati et al.]{Matteo Fossati$^{1,2}$\thanks{E-mail: mfossati@mpe.mpg.de}, Michele Fumagalli$^{3}$, Alessandro Boselli$^4$, Giuseppe Gavazzi$^5$, 
  \newauthor Ming Sun$^6$, David J. Wilman$^{1,2}$\\
  $^{1}$Universit{\"a}ts-Sternwarte M{\"u}nchen, Scheinerstrasse 1, 81679 M{\"unchen}, Germany \\
  $^{2}$Max-Planck-Institut f{\"u}r Extraterrestrische Physik, Giessenbachstrasse, 
  85748 Garching, Germany\\
  $^{3}$Institute for Computational Cosmology and Centre for Extragalactic Astronomy, 
  Department of Physics, Durham University, South Road, Durham, DH1 3LE, UK \\
  $^{4}$Aix Marseille Universit{\'e}, CNRS, 
  LAM (Laboratoire d' Astrophysique de Marseille) UMR 7326, 13388, Marseille, France \\
  $^{5}$Universit{\'a} di Milano-Bicocca, piazza della Scienza 3, 20100, Milano, Italy \\
  $^{6}$Department of Physics, University of Alabama in Huntsville, Huntsville, AL 35899, USA \\
}

\begin{document}

\date{Accepted 2015 October 14. Received 2015 October 13; in original form 2015 August 17}

\pagerange{\pageref{firstpage}--\pageref{lastpage}} \pubyear{2015}

\maketitle

\label{firstpage}

\begin{abstract}
We present a study of the physical properties of the disc and tail of \eso, a 
galaxy suffering from extreme ram-pressure stripping during its infall into the 
Norma cluster. With sensitive and spatially-resolved MUSE spectroscopy, 
we analyse the emission line diagnostics in the tail of \eso, 
finding high values of [NII]/\ha\ and [OI]/\ha\ 
that are suggestive of the presence of shocks in turbulent gas.
However, the observed line ratios are not as strong as commonly seen in pure shock 
heating models, suggesting that other emission mechanisms may contribute to the 
observed emission. Indeed, part of the observed emission, particularly at close 
separations from the galaxy disc, may originate from recombination of photoionized gas stripped 
from the main body of \eso. We also identify a large number of bright compact knots 
within in the tail, with line ratios characteristic of HII regions. 
These HII regions, despite residing in a stripped gas tail,
have quite typical line ratios, densities, temperatures, and metallicity ($\sim0.7$ solar). 
The majority of these HII regions are embedded within diffuse gas from the tail, 
which is dynamically cool ($\sigma \sim 25-50$ \kms). This fact, together with a lack 
of appreciable gradients in age and metallicity, suggests that these HII regions formed
in situ.
While our analysis represents a first attempt to characterise the rich physics of the 
\eso\ tail, future work is needed to address the importance of other mechanisms, 
such as thermal conduction and magneto hydrodynamic waves, in powering the emission in the tail. 
\end{abstract}


\section{Introduction}

It has long been known that galaxy colour and morphology correlate both with local environment 
and stellar mass \citep{dressler80,kennicutt83,postman84,bell04,peng10}, an empirical manifestation 
that both internal and external physical processes play an important role in shaping the 
star formation of galaxies at all cosmic ages \citep[see e.g.][]{boselli06,blanton09}. 
Many studies point to internal processes, which often correlate with stellar or bulge mass,  
as key drivers for quenching star formation activity at all densities \citep{cowie96,gavazzi96,mendel13, wilman13, lang14}. 
However, the advent of large scale surveys has also highlighted and strengthened earlier findings that
environmental effects in rich clusters and groups are key drivers for the quenching of star formation
\citep{balogh04,baldry06,weinmann06,boselli06,boselli14,gavazzi10,wilman10,mok13}.

In this respect, rich galaxy clusters are the best laboratories for investigating the processes 
that are responsible for quenching star formation, due to the combination of strong potential 
wells, high number density of galaxies, and the presence of the intracluster medium (ICM),
a hot and dense plasma filling the intergalactic space. Several mechanisms have been proposed
as the culprit for galaxy transformation within clusters, including tidal interactions between 
galaxies (sometimes called galaxy harassment) or with the cluster potential itself 
\citep{byrd90,henriksen96,moore96}, galaxy strangulation \citep{larson80}, and ram pressure 
stripping \citep{gunn72}.

Although more than one of these mechanisms likely play a role in the evolution of cluster 
galaxies at one time, a few pieces of observational evidence consistently point to 
ram pressure stripping as a widespread process, and one of crucial importance 
for the transformation of low mass galaxies that are being accreted on to 
clusters at recent times \citep{boselli08}. 
Observationally, the fingerprint of ram pressure is the removal of the galaxy gas content
without inducing strong disturbances in the older stellar populations.  
Indeed, large radio surveys have found that the HI content of cluster (and to some extent group)
galaxies is being reduced with respect to their field counterparts 
\citep{giovanelli85,solanes01,cortese11,fabello12,catinella13,gavazzi13,jaffe15}, an effect that has been 
also recently observed in the molecular phase \citep{fumagalli09, boselli14b}. 
This gas depletion, in turn, induces truncated profiles in 
young stellar populations compared to old ones \citep{koopmann06,boselli06b,cortese12,fossati13}. 

Besides understanding the influence of ram pressure on cluster galaxies as a 
population, a detailed look at how this mechanism operates on small scales within 
individual galaxies is becoming a pressing issue, so as to gain insight into how
gas removal occurs as a function of parameters such as galaxy mass or 
distance from the cluster centre. Also poorly understood is the fate of the stripped 
gas, which could mix with the hotter ICM or cool again to form stars, thus contributing 
to the intracluster light. Fortunately, these detailed investigations are 
now within reach, thanks to the advent of integral field spectrographs at 
large telescopes. 

Ram pressure has been caught in the act by searching for UV and \ha\ tails downstream 
of galaxies in fast motion through massive clusters like Coma or Virgo 
\citep{gavazzi01, yoshida02, cortese06, kenney08, smith10, yagi10, fossati12}. 
The amount of star formation in stripped tails has been the subject of several analyses, 
both from the observational \citep{yagi10, smith10, fummattia11, arrigoni12, boissier12, fossati12} and 
theoretical point of view \citep[e.g.][]{kapferer09, tonnesen12}.
However, the efficiency of star formation and its dependence on the gas properties of the 
tails remain poorly understood. 

A posterchild for studies of ram pressure stripping is \eso, a spiral galaxy near to the 
core of the Norma cluster. The Norma cluster, also known as A3627, has a dynamical mass 
of $M_{\rm dyn} \sim 10^{15} \rm{M\odot}$ at a redshift $z_{\rm cl} = 0.01625 \pm 0.00018$ 
\citep{woudt08, nishino12}. \eso\ is located at a projected distance of only $\sim 267$ kpc 
from the central cluster galaxy, and features a wealth of multiwavelength observations 
ranging from X-ray to radio wavelengths \citep{sun06,sun07,sun10,sivan10,jac14}.

All these observations clearly show that \eso\ is suffering from extreme ram pressure stripping, 
arguably during its first infall into the cluster environment. Indeed, while 
\eso\ can be classified as a normal spiral galaxy from its stellar continuum, 
X-ray and \ha\ observations reveal the presence of a double tail that extends for $\sim 80$ kpc 
behind the galaxy disc, in the opposite direction from the cluster 
centre \citep{sun06, sun07, sun10}. These tails are believed to originate from the hydrodynamic 
interaction between the hot ICM and the cold interstellar medium (ISM), 
which is removed from the galactic disc \citep{sun07, jac14, pap1}. 
Although stripped tails have been extensively studied in imaging in various nearby clusters, 
the number of spectroscopic studies in the optical has been limited in the past 
\citep[e.g.][]{yagi07, yoshida12, merluzzi13}, due to the extended nature and the low surface brightness 
of the emitting regions.

To overcome the lack of extended spectroscopic follow-up of this galaxy, 
we have collected integral field spectroscopic observations of \eso\
following the deployment of Multi Unit Spectroscopic Explorer \citep[MUSE;][]{bacon10} at the 
ESO Very large Telescope. These observations have been presented in the first paper of this 
series \citep[][hereafter paper I]{pap1}.  The combination of high efficiency, spectral resolution 
and large field of view of MUSE have allowed us to map the kinematics of the stripped tail for the 
first time. In paper I, we found that the stripped gas retains the imprint of the disc rotational 
velocity up to $\sim20$ kpc downstream, with low velocity dispersion.
Further on along the tail, the gas velocity dispersion increases up to $\sim 100$ \kms. 
Moreover, we observed an ordered velocity field for the stellar disc, which convincingly 
points to ram pressure as the mechanism for gas removal. 

In this paper, we extend our previous analysis by investigating the physical properties of the 
ionized gas and the star formation in the tail of \eso\ by means of multiple emission line
diagnostics. The structure of this paper is as follows. After discussing an improved data reduction 
scheme for MUSE observations (Section 2) and the data analysis technique (Section 3), we present 
a study of the emission line ratios as diagnostic of the physical conditions of the gas in the 
tail (Section 4). Lastly, we focus on several HII regions detected throughout the tail,
and briefly discuss possible formation mechanisms.
Throughout this paper, we adopt a standard $\Lambda$CDM cosmology with $H_0 = 69.7$ 
and $\Omega_{\rm m} = 0.236$ \citep{hinshaw13}. With the adopted cosmology, 
1 arcsec corresponds to 0.32 kpc at the distance of \eso.

\section{Observations and data reduction}

\eso\ was observed with MUSE during the science verification programme 60.A-9349(A) on 2014 June
21, ionized. The details of the observational setup are presented in paper I, and they 
are only briefly summarised here. Observations were acquired in photometric conditions and 
under good seeing ($0.7"-0.8"$), using the Wide Field Mode with nominal wavelength coverage 
(4800-9300 \AA). The spectral resolution is $\sim 50$ \kms\ and roughly constant across the 
wavelength range.  The raw data have been processed with the MUSE data reduction pipeline 
(v 0.18.1), which we supplemented with custom-made IDL codes to improve the quality of the 
illumination correction and of the sky subtraction. As discussed in paper I, we first apply to each 
exposure an illumination correction, which we obtain by comparing the brightness of multiple sky 
lines to compensate for variations in the response of the different MUSE spectrographs. 

\begin{figure} 
\includegraphics[scale=0.47]{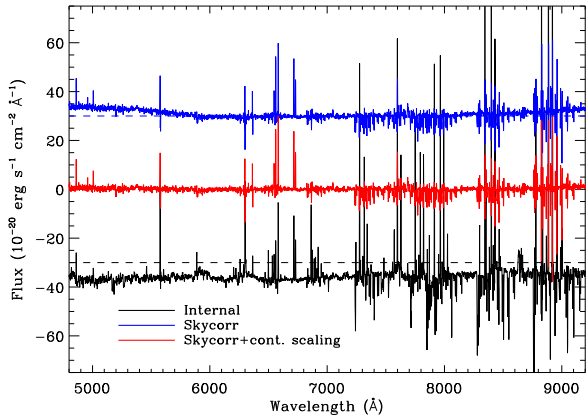}
\caption{Sky residuals recorded in a 10-pixel square aperture within one science exposure that has 
  been processed with different algorithm for sky subtraction. The black line shows the result of the 
  MUSE pipeline in this crowded field. The blue line is obtained by subtracting a sky model derived 
  from a second sky exposure together with the scaling computed with {\sc skycorr}. Both spectra are 
  visibly offset from the zero level, marked by the dashed lines. The red line shows the sky residual 
  in the data cube, after including scaling in both the airglow lines and the continuum to match the 
  level recorded in the science exposure. Large sky line residuals do not appear in the final 
  processed cube (the group of lines at $\sim 6500$\AA, whose residuals do not change in the 
  three spectra, are in fact emission lines from the Galaxy).}
  \label{skysubcompare}
\end{figure}

After this step, we perform sky subtraction using in-house codes, as the very crowded nature 
of our field requires additional processing to avoid large negative residuals, as shown in Figure 
\ref{skysubcompare}. In this work, we further improve on the method adopted in paper I, so as 
to maximise the quality of the sky subtraction across the entire wavelength range of the data cube. 
In particular, because a significant fraction of our field is filled by the emission lines from the 
tail of \eso\ or by the continuum emission from the stellar disc and halo, we use exposures of empty 
sky regions to avoid the inclusion of spurious signal arising from the galaxy itself in the sky model. 
Specifically, we use a sky exposure of 60s, acquired $\sim 1$ hour apart from our science observations 
by the programme 60.A-9303(A).
 
As a first step, we generate a supersampled sky spectrum (hereafter the sky model) after processing 
the pixel table of the sky exposure with the same pipeline recipes used for our science exposures. 
This sky model, however, cannot be directly subtracted from the science exposures because of the 
time dependency of both the airglow line and continuum fluxes. To correct for this variation, 
we exploit the ESO tool {\sc skycorr} \citep{nol+14} to compute a physically-motivated scaling of 
the airglow lines in the sky model to match the levels recorded in a model spectrum extracted 
from each of the science exposures. Empirically, we find that {\sc skycorr} produces line models 
which yield small residuals after subtraction, as shown in Figure \ref{skysubcompare}.
At this stage, however, {\sc skycorr} does not attempt any scaling of the sky continuum emission, 
mainly because the continuum emission from the sources cannot be easily separated from the sky 
continuum level. Thus, to set a zero-point for the sky continuum emission, we take advantage of the 
large field of view of MUSE which contains regions that are not affected by the galaxy's 
continuum flux. After subtracting the rescaled sky model from the science exposures, any residual 
deviation from a zero continuum level in these ``empty'' regions arises because of the different 
sky continuum level in the science and the sky exposures (see Figure \ref{skysubcompare}). To remove 
this last signature, we fit two third-degree polynomials in regions free from sky lines above and 
below 6200 \AA, that is the wavelength where the sky continuum reaches its peak. Finally, we use 
this polynomial fit to correct for the residuals due to varying sky continuum emission 
(Figure \ref{skysubcompare}). During these steps, the sky model is subtracted from each pixel of 
the data cube by interpolating on the wavelength axis with a spline function. The appropriate 
uncertainties are added to the variance stored in the data cube (the ``stat'' extension). 

After sky subtraction, we use the MUSE pipeline to combine the four individual exposures and 
reconstruct the final data cube on a regular grid of 0.2 arcsec in the spatial direction and 
1.25 \AA\ in the spectral direction. As a last step, we apply the heliocentric correction and 
we correct the observed flux as a function of wavelength to account for a substantial amount 
of Galactic extinction in the direction of \eso. This correction is computed using the colour 
excess from the dust map by \citet{sch98} with the recalibration of \citet{sch11}. We also 
assume a Galaxy extinction curve from \citet{fit99}.

In paper I, we have verified the accuracy of the flux calibration by comparing the \ha\ 
fluxes of selected HII regions to the measurements reported by \citet{sun07}. We further 
verify that the flux calibration is consistent across the entire wavelength range by 
checking that the spectra of stars in the field are well fitted by blackbody functions. 
Moreover, we compare the flux ratios of several emission lines derived in this work to those 
from \citet{sun10} for their HII regions ELO2 and ELO7, finding agreement within 5\% for all 
the flux ratios compared. 

\section{Emission line measurements} 

To characterise the physical properties of the gas in the disc and the tail, we extract flux 
maps of the emission lines listed in Table \ref{emlines}. In this table, we also provide for each 
line an estimate of the noise, quantified as the $3\sigma$ limit per spaxel in the unsmoothed MUSE 
data cube. As in paper I, we assume a systemic redshift $z_{\rm sys} = 0.01555 \pm 0.00015$ for \eso. 
Maps of the line fluxes are obtained from the 
reduced datacube using the {\sc idl} custom code {\sc kubeviz}.  Before the fit, the datacube 
is median filtered in the spatial direction with a kernel of $15 \times 15$ pixels (corresponding to 
$3''$ or 0.95 kpc at the distance of \eso) to increase the signal-to-noise ($S/N$) per pixel without 
compromising the spatial resolution of the data. No spectral smoothing is performed. 

In the galaxy disc, Balmer emission lines (\ha\ and \hb) are contaminated by stellar absorption. 
Direct fitting of the emission lines would thus underestimate the fluxes. To solve this problem, we first 
model and subtract the underlying galaxy stellar continuum using the {\sc gandalf} code \citep{sarzi06} for
spaxels in a rectangular area that covers the stellar disc (see figure 5 in paper I) 
and with S/N$>5$ per velocity channel in the continuum. 
{\sc gandalf} works in combination with the Penalized Pixel-Fitting code \citep{cappellari04} to simultaneously 
model the stellar continuum and the emission lines in individual spaxels. The stellar continuum is 
modelled with a superposition of stellar templates convolved by the stellar line-of-sight velocity distribution, 
whereas the gas emission and kinematics are derived assuming a Gaussian line profile. To construct the stellar 
templates, we use the MILES library \citep{vzadekis10}. 

\begin{figure*} 
\includegraphics[scale=0.90]{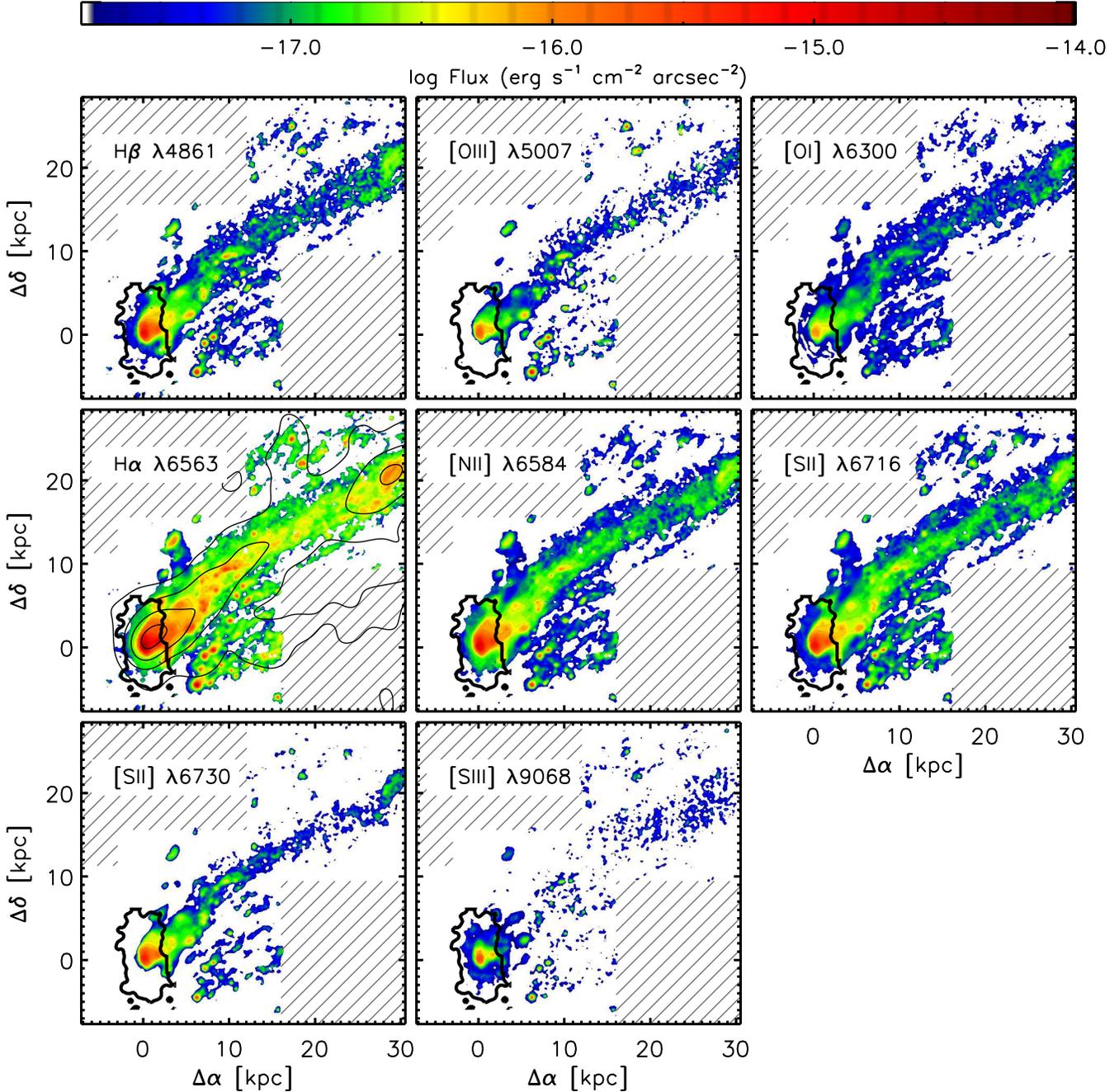}
\caption{Emission line maps of \eso. The panels are sorted by the wavelength of the line from left 
to right and top to bottom. The black solid contour marks the isophote of the galactic disk at 22 mag arcsec$^{-2}$ 
measured on an $r$-band image obtained from the datacube. The thinner solid contours in the \ha \ panel show
the location of the X-ray emitting gas from \textit{Chandra} observations \citep{sun10}, highlighting the primary and secondary tails.
Areas not covered by MUSE 
observations are shaded in grey.}\label{galleryemission}
\end{figure*}

Having obtained a data cube free from underlying stellar absorption, we use {\sc kubeviz} to fit the emission 
lines. This code uses ``linesets'', defined as groups of lines that are fitted simultaneously. Each lineset 
(e.g. \ha\ and [NII] $\lambda \lambda 6548,6584$) is described by a combination of 1D Gaussian 
functions where the relative velocity separation of the lines is kept fixed according to the wavelengths listed 
in Table \ref{emlines}. To facilitate the convergence of the fit for the faintest lines, we impose a prior on the 
velocity and the intrinsic line width $\sigma$ of each lineset, which is fixed to that obtained fitting the \ha\ 
and [NII] lines. We have explicitly verified on the master spectra described in Section \ref{linediag} that this 
assumption holds well both in the disc and across the tail. Furthermore, the flux ratio of the two [NII] and [OIII] 
lines are kept constant in the fit to the ratios in \citet{storey00}. The continuum level is evaluated inside 
two symmetric windows around each lineset. Figure \ref{galleryemission} shows emission line maps obtained from 
our fits of the datacube at selected wavelengths. 
We note that, at the redshift of \eso, [SII] $\lambda6731$ falls onto a bright sky line at $\lambda6834$ and 
thus the strong residual hampers a reliable estimate of the flux for this line (see Figure \ref{reg:spect}). 
Therefore, we are forced to exclude [SII] from most of the following analysis.

During the fit, {\sc kubeviz} takes into account the noise from the ``stat'' data cube, thus optimally suppressing 
sky line residuals. However, the adopted variance underestimates the real error, most notably because it does not 
account for correlated noise introduced by resampling and smoothing. We therefore renormalise the final errors 
on the line fluxes assuming a $\chi^2 =1$ per degree of freedom. In the end, as in paper I, 
we mask spaxels where the $S/N$ in the \ha\ line is $<3$. Further masking is applied to the spaxels for which 
the line centroids or the line widths are extreme outliers compared to the median value of their distributions, 
or the errors on these quantities exceed 50 \kms.

\begin{table}
\caption{Emission lines considered in this study}\label{emlines} 
\centering
\begin{tabular}{l c c}
\hline
\hline
Line&$\lambda_{\rm r}$ & $\mu_{\rm min}$                    \\
    &     (\AA)        & ($\rm erg~s^{-1}~cm^{-2}~A^{-1}~arcsec^{-2}$)  \\
\hline  
  H$\beta$         &  4861.33  & $8.2 - 11.7 \times 10^{-18} $ \\ %
  \ion{[O}{III}]   &  4958.91  & $7.7 - 10.7   \times 10^{-18}  $ \\ %
  \ion{[O}{III}]   &  5006.84  &  $7.5 - 10.5  \times 10^{-18}  $ \\ %
 \ion{[N}{II}]	   &  5754.64  & $ 6.2 - 8.7  \times 10^{-18}  $ \\ %
  \ion{[O}{I}]	   &  6300.30  &  $4.6 - 6.3   \times 10^{-18} $ \\ %
  \ion{[O}{I}]	   &  6363.78  &  $5.0 - 6.7  \times 10^{-18}  $ \\ %
  \ion{[N}{II}]    &  6548.05  &  $4.1 - 5.6  \times 10^{-18}  $ \\ %
  H$\alpha$	   &  6562.82  & $ 4.1 - 5.6  \times 10^{-18}  $ \\ %
  \ion{[N}{II}]    &  6583.45  &  $4.1 - 5.5 \times 10^{-18}   $ \\ %
  \ion{[S}{II}]    &  6716.44  &  $4.9 - 6.4  \times 10^{-18}  $ \\ %
  \ion{[S}{II}]    &  6730.81  &  $5.0 - 6.6  \times 10^{-18}  $ \\ %
  \ion{[S}{III}]   &  9068.60  &  $4.8 - 6.9 \times 10^{-18}   $ \\ %
\hline    
\end{tabular}
\flushleft{The columns of the table are: (1) name of the emission line; (2) wavelength in air; 
(3) characteristic surface brightness limit at $3\sigma$ C.L. computed in each spaxel after correcting for Galactic extinction.
The two values refer to pointing A and B respectively.}
\end{table}

\section{Emission line diagnostics} \label{linediag}

Several physical processes, including photoionisation, shocks, and thermal energy
in mixing layers, are likely to contribute to the emission lines we see in the spectra of the 
main body of \eso\ and its tail. In this work we primarily focus on the photoionisation
and shocks. In a forthcoming paper, new observations (095.A-0512(A)) covering the 
full extent of the primary and the secondary tails will be used in combination with 
multiwavelength data to study the interaction between the warm ionized gas in the 
tail and the hot ICM. 

\subsection{Line ratio maps} \label{lineratiomaps}

In this section, we exploit the integral field capabilities of MUSE to investigate the spatial variations
of physical properties inferred from strong emission line diagnostics computed for each pixel.
Next, we will coadd the spectra of several characteristic regions of the \eso \ system
to achieve higher $S/N$ in the fainter lines. 

\begin{figure*} 
\includegraphics[scale=0.90]{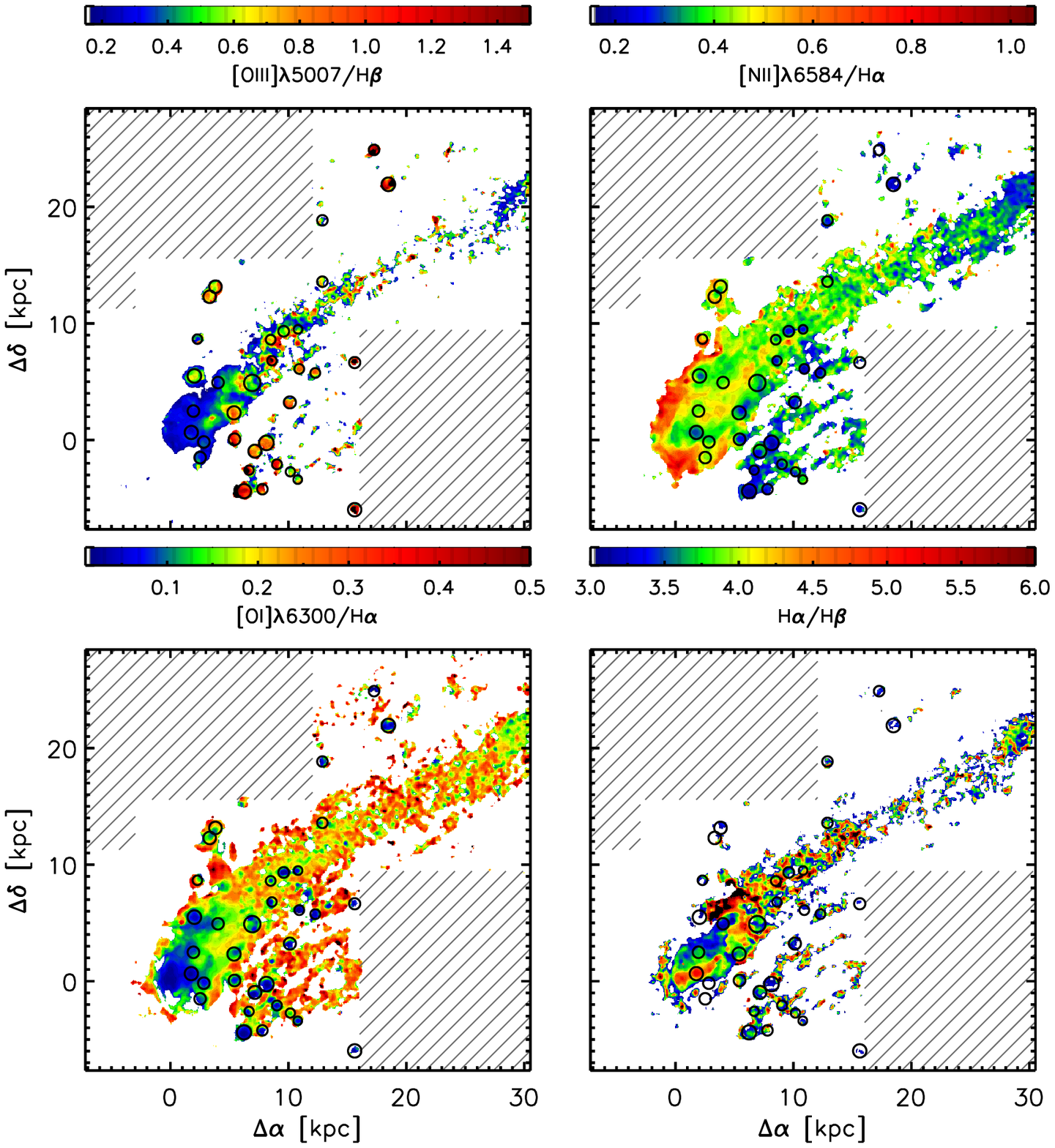}
\caption{Maps of the [OIII]$\lambda 5007/$\hb \ , [NII]$\lambda 6584/$\ha, [OI]$\lambda 6300/$\ha, and \ha/\hb\ 
line ratios Areas not covered by MUSE observations are shaded in grey. The compact knots
associated with HII regions along the tail are highlighted with black circles in all the panels.}
\label{physics2D}
\end{figure*}

The ratio of collisionally excited lines like [OIII]$\lambda5007$, [NII]$\lambda6584$ to hydrogen
recombination lines (\ha, \hb ) is traditionally used as indicator of the ionisation properties of the 
gas. More specifically, the combination of pairs of these line ratios is used to construct BPT diagrams 
\citep{baldwin81}, so as to distinguish different sources of ionisation, e.g. photoionization in star forming 
galaxies or active galactic nuclei, and shocks. These line ratios are typically chosen 
such that both lines fall inside a small window of wavelength to minimise the effects of dust extinction. 

The top panels of Figure \ref{physics2D} show 2D maps of the [OIII]$\lambda 5007/$\hb, 
[NII]$\lambda 6584/$\ha, and [OI]$\lambda 6300/$\ha \ ratios, computed for all the spaxels 
in which both of the lines of interest are detected with $S/N > 3$. Significant spatial variations in 
these three line ratios can be seen in the data. Most of the large scale trends visible in this figure are real, 
but we caution that two effects should be taken into account when interpreting the maps.
First, in the outer tail, the depth of the observations is shallower due to shorter exposure time. 
Second, in the front region, the smoothing applied to the datacube can artificially enhance 
discontinuities in line ratios and affect their gradients.

Inspecting the H$\alpha$ map (left central panel in Figure \ref{galleryemission}), compact knots along the tail 
can be identified as high surface brightness regions. To isolate these knots, we 
run {\sc SExtractor} v2.19.5 \citep{bertin96} on the \ha \ map.  Among the detected objects (S/N $>5$) we selected 
those with a classification consistent with point-like sources (CLASS\_STAR $> 0.9$) and low ellipticity 
($e<0.2$). 33 such knots are identified in our data. Out of these 27 correspond to the
HII regions analysed in \citet{sun07, sun10}.  The remaining six regions 
lie close to foreground stars, making their selection more difficult in the narrow band imaging data 
of \citet{sun07}. In Figure \ref{physics2D}, we have highlighted the position the compact knots 
with black circles. These regions are mainly characterised by high values of [OIII]$\lambda 5007/$\hb\ ($\gtrsim 0.7$)
as well as low values of [NII]$\lambda 6584/$\ha\ ($\lesssim 0.3$) and [OI]$\lambda 6300/$\ha\ ($\lesssim 0.1$) 
ratios. According to the BPT classification (see also Figure \ref{BPTdiag}), all these regions occupy the parameter space
populated by HII regions photoionized by OB stars. 

Focusing next on the galaxy nucleus, at the origin of the coordinate system in Figure \ref{physics2D},
we see modest ratios for [OIII]$\lambda 5007/$\hb, [NII]$\lambda 6584/$\ha\ and [OI]$\lambda 6300/$\ha,
which is again consistent with ionisation from a soft spectral energy distribution.
Thus, in agreement with the X-ray analysis of \citet{sun10}, we conclude that 
the nuclear emission of \eso\ is powered by star formation activity and that this
galaxy does not host a strong AGN.

Looking at the galaxy front next, where the ISM collides with the hot plasma from the cluster, 
we see enhanced [NII]$\lambda 6584/$\ha\ ($\gtrsim 0.6$) as well as low [OIII]$\lambda 5007/$\hb\ ($\lesssim 0.3$), 
with a hint of elevated [OI]$\lambda 6300/$\ha. These ratios
are suggestive of shocks playing a role in the gas excitation \citep{allen08, rich11}.  
Throughout the primary tail [NII]$\lambda 6584/$\ha\ remains elevated ($\sim0.4$) 
decreasing only towards the end of the tail, in proximity of the extended \ha\ blob that 
is known to harbour a large molecular complex \citep{jac14}.
We also find enhanced [OI]$\lambda 6300/$\ha\ ($\gtrsim 0.2$) in the tail, while 
the [OI] line is weaker in the galactic disc. 
This piece of evidence, combined with the high gas turbulence measured in paper I, 
points to shock heating as an important mechanism to excite the gas in the tail. 

Differently from the primary tail, the southern secondary tail shows a strong [OI] emission,
but a lower [NII]$\lambda 6584/$\ha, which is on average $\sim 0.2$ after removing the HII regions.
These line ratios are more difficult to reconcile with emission lines from ordinary shock models.
We note however that the presence of several bright HII regions embedded in this tail can 
produce mixed types of spectra, with a superposition of emission lines that originate 
from different emission mechanisms in distinct gas phases.
More generally, we are witnessing a complex interaction between multiple gas phases as the 
warm ISM from the galaxy, ionized by past and ongoing star formation, mixes and interacts with 
the hot ICM. Thus, photoionisation from stars,  shocks, and even thermal conduction and magnetohydodynamic (MHD) waves 
\citep[see e.g.][]{ferland08, ferland09}, are likely contributing to a varying degree to the peculiar 
line ratios seen in \eso.

We conclude this section by briefly commenting on the \ha/\hb\ ratio, shown in the forth panel of 
Figure \ref{physics2D}. This ratio is commonly used in the low density limit to infer
the dust extinction, assuming an intrinsic ratio of 2.86 from case B recombination
at $T=10^4~$K \citep{ost89}. As the observed emission, especially in the tail, 
may not simply arise from photoionized gas at the temperature of typical HII regions 
(see Sect. \ref{sec:compositespec}), we refrain from deriving a spatially-resolved map for the extinction. We only note 
that ratios are generally \ha/\hb$\gtrsim 3$, which is suggestive of the presence of dust both in 
the disc and tail of \eso. \citet{cortese10a, cortese10b} indeed observed dust stripping from galaxies
in cluster environments, supporting the idea that gas and dust are subject to similar perturbations.

\subsection{Composite spectra}\label{sec:compositespec}
Having examined qualitatively the spatial variation of the different line ratios,
in this section we exploit the ability to rebin IFU data in regions of interest. We 
generate high S/N composite spectra for a quantitative analysis of even
weak emission lines. We focus on five characteristic regions,
which exhibit different line ratios in Figure \ref{physics2D}.
These are: 
1) the galaxy disc, including the nucleus; 
2) the front region, defined as the area with enhanced [NII]/\ha\ ratio in the south-east of the galaxy disc; 
3) the compact knots along the tail, which we have identified as HII regions; 
4) the diffuse gas within the primary tail; 
5) the H$\alpha$ ``blob'' in the outer tail. 

\begin{figure} 
\includegraphics[scale=0.45]{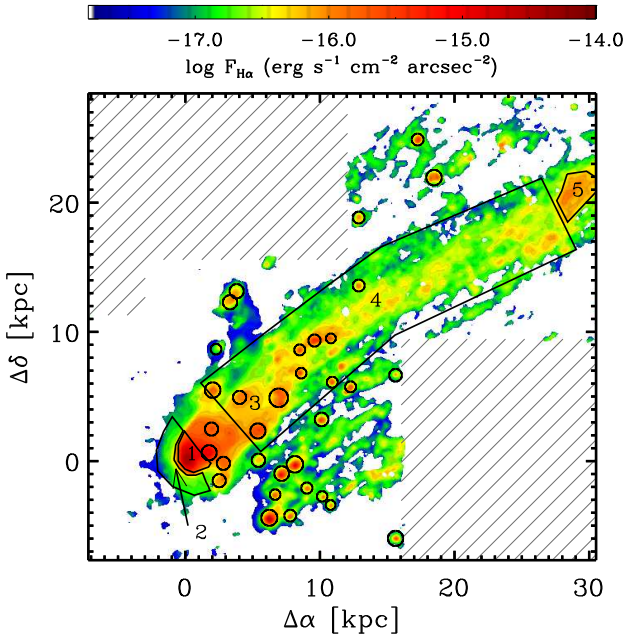}
\caption{\ha\ map of \eso. Black apertures represent selected regions used to create composite 
  spectra as described in the text.
  Specifically, circular apertures identify HII regions (3), while polygons
  identify (from left to right): the front region (2); the main body of the galaxy (1); 
  the diffuse gas within the tail (4); the blob in the outer tail (5). 
  Areas not covered by MUSE observations are shaded in grey. }\label{reg:img}
\end{figure}

\begin{figure*} 
\includegraphics[scale=0.70]{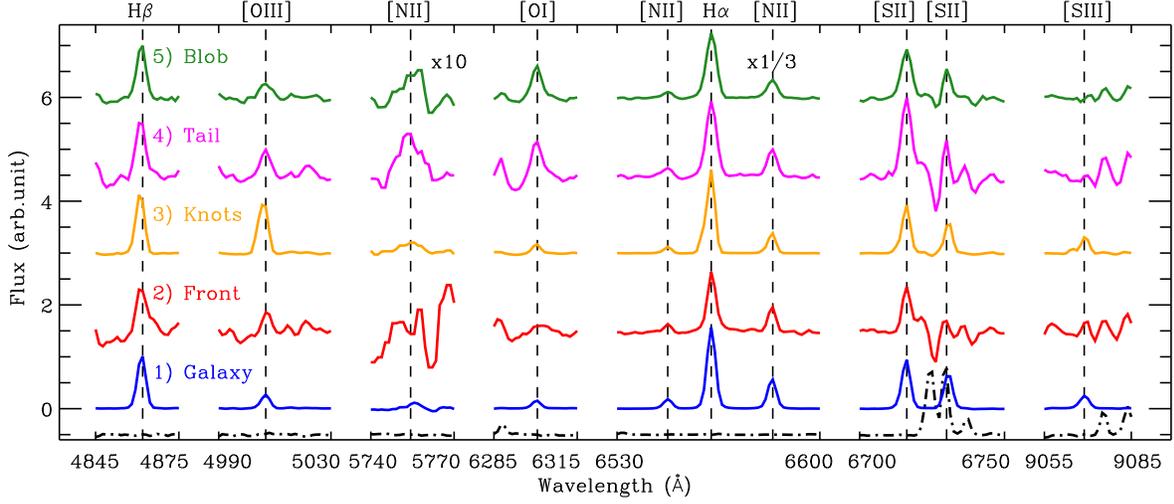}
\caption{Coadded rest frame spectra of the five regions of \eso\ shown in figure \ref{reg:img}. 
The spectra are normalised to the \hb\ intensity and shifted along the vertical axis 
to facilitate the comparison. Only the emission lines relevant to this study are shown. In case of 
[OIII], and [OI] doublets only the strongest line is shown.
For visualisation purposes, the \ha\ and [NII] complex have been scaled by a factor of three, 
while the [NII]$\lambda5755$ is enhanced a factor of 10. The lowermost spectrum (dash-dotted line) 
is a night sky spectrum (not in scale) extracted from the datacube. }
\label{reg:spect}
\end{figure*}

The locations of these different regions are shown in Figure \ref{reg:img}, while Figure \ref{reg:spect}
shows the composite spectra obtained by co-adding all spaxels in the selected apertures with 
a mean. Before generating the composite spectra, we 
remove the velocity offsets due to the systemic redshift differences that arise because of the galaxy 
rotation. For these corrections we rely on the gas velocity field derived on the new reduction with the 
methods described in paper I. 
For the regions 1 and 2, we further use the {\sc gandalf} code to remove the 
stellar continuum spectrum from the composite spectrum, as described above. The composite spectrum 
of the HII regions is instead obtained by averaging the individual spectra with inverse variance 
weights derived from the bootstrap noise spectra described below.

Figure \ref{reg:spect} reveals that the wavelength at which all the 
emission lines peak is consistent with the galaxy's redshift regardless of the transition
under consideration. The largest offsets occur predominantly in the regions where the lines are fainter,
but these are generally consistent with zero given the uncertainties on the line centroid.
This empirical evidence justifies our previous assumption to peg the centroid of different emission lines
to the one of \ha\ when deriving the flux maps in the full data cube. 

Furthermore, as already discussed above, this figure readily shows that the flux ratios 
change as a function of the spatial position in the disc of the galaxy and in the tail,
which is suggestive of various excitation mechanisms at work in different regions. 
For instance, the front region has the highest [NII] to \ha\ ratio, while [OIII]$\lambda5007$ 
is most strongly detected in the bright \ha\ knots along the tail.
Furthermore, [OI]$\lambda6300$ emission is visibly stronger throughout the tail than in 
the main galaxy body. 

This visual assessment of the data is more quantitatively supported by the integrated fluxes, listed in Table 
\ref{tab:ratios}. The uncertainties on the line fluxes are computed from 100 bootstrap realisations of 
the median spectra. More specifically, the fitting procedure is repeated for each bootstrap realisation, and
we assume the 1$\sigma$ confidence level of the measured line fluxes as the formal uncertainty. This approach,
however, does not take into account the correlated noise contribution, which is estimated to be 
a factor $\sim 2$ for IFU observations where a similar number of pixels is combined \citep[e.g.][]{forster09}.  
For undetected lines, we quote the 3$\sigma$ confidence level obtained from the standard deviation of the 
continuum values on both sides of each line.

In the next section, we will attempt to characterise the density, temperature, and 
metallicity of the gas, by means of emission line diagnostics leveraging
a combination of spatially resolved maps and composite spectra.

\subsection{Properties of the diffuse gaseous tails}

\subsubsection{Density}

The electron density in the ionized gas can be measured, for instance, through collisional deexcitation
in doublets of the same ion from different levels with comparable excitation energy. 
One such doublet is [SII] $\lambda6716/\lambda6731$. Unfortunately, the intensity ratio is only weakly 
sensitive to the density for $n_e < 10^2~\rm cm^{-3}$, where collisional excitation is generally followed
by photon emission. Moreover, in the case of \eso, one of the [SII] lines falls onto a bright skyline, 
severely compromising the derived fluxes especially in the low surface brightness regions of the tail. 
For these reasons, we are forced to resort to other estimators for the gas density. 

A crude order of magnitude estimate for the electron density can be obtained from the observed \ha\ 
luminosity, together with an estimate of the volume. 
More specifically, in equilibrium, the \ha \ luminosity in a given spatial region can be written as
\begin{equation}
{L_{\rm H\alpha} = n_e n_p \alpha_{\rm H\alpha}^{\rm eff} V f h \nu_{\rm H\alpha} }
\label{eq:Lha}
\end{equation}
where $n_e$ is the number density of electrons, $n_p$ is the number density of protons (hydrogen ions), 
$\alpha_{\rm H\alpha}^{\rm eff}$ is the \ha \ effective recombination coefficient, $V$ is the volume of the emitting region, 
$f$ is the filling factor, $h$ is the Planck's constant and $\nu_{\rm H\alpha}$ is the frequency of the 
\ha\ transition. 

\begin{figure} 
\includegraphics[scale=0.45]{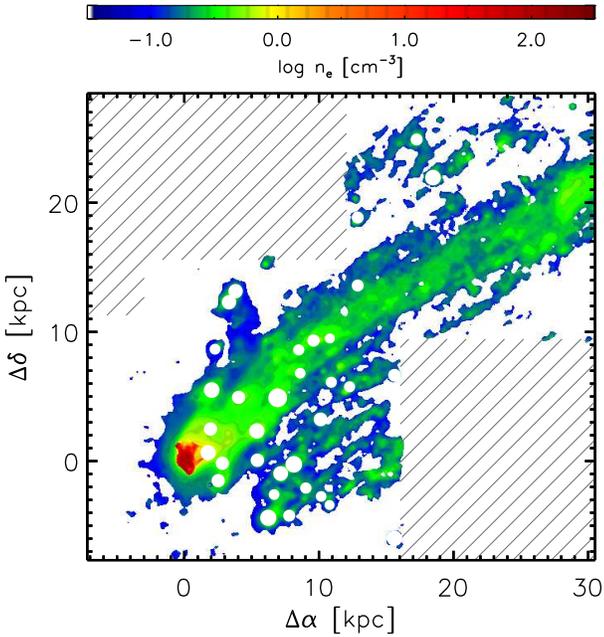}
\caption{Map of the mean election density obtained from the \ha \  surface brightness and a cylindrical volume 
from the tail. A filling factor $f=0.05$ is assumed. Areas not covered by MUSE observations 
are shaded in grey. The HII regions along the tail are excluded from the map.}
\label{density2D}
\end{figure}

For the geometry of the tail, we assume a cylinder with line-of-sight depth 
equal to the diameter in the plane of the sky. The diameter, as measured on the \ha\ image, is 
$D \sim0.3$ arcmin throughout the entire length of the tail, or about $D \sim5.5$ kpc 
at the distance of \eso. This value is $\sim 50\%$ larger than the diameter used by \citet{sun07},
as the superior depth of the MUSE observations allows us to also detect the 
lower surface brightness components of the tail. In our calculation, we consider single spaxel with area 
$A$ as individual regions. The filling factor is instead unconstrained. However, there is indication 
that the warm ionized gas in the tail can be very clumpy, as shown for instance 
in hydrodynamic simulations of ram-pressure stripping \citep[e.g.][]{tonnesen11}.
Thus, we assume $f=0.05$, as in \citet{sun07}. Finally, we assume that the gas is fully ionized 
($n_e = n_p$) and $\alpha_{\rm H\alpha}^{\rm eff} =1.17 \times 10^{-13}~\rm cm^3~s^{-1}$ \citep{ost06}. 
As previously noted, the excitation mechanisms of the tail are unknown and 
the value of the recombination coefficient is uncertain. Moreover, the assumption of a cylindrical volume
implies that the line-of-sight depth of the cylinder depends on the spatial position. However our 
calculation aims to provide only a order of magnitude estimate for the gas density and other quantities, 
such as the filling factor, are likely to dominate the error budget in this calculation.
Under these assumptions, manipulating equation \ref{eq:Lha} we obtain
\begin{equation}
n_e  = \sqrt{\frac{L_{\rm H\alpha}}{\alpha_{\rm H\alpha}^{\rm eff} A D f h \nu_{\rm H\alpha} }}
\label{eq:densitytail}
\end{equation}

Figure \ref{density2D} shows the resulting map for the gas density obtained following equation \ref{eq:densitytail}, 
where HII regions have been excluded from the calculation.
From the map, we find that the average density in the tail, 
although uncertain, is of the order of $0.3~\rm cm^{-3}$,  in agreement with the estimate of \citet{sun07}.
Very similar densities are found by \citet{yagi07} and \citet{fossati12} in ram pressure stripped tails in
the Coma cluster, once a common filling factor is assumed.
As a cross check, we also examine the composite spectrum of the tail, 
where the improved S/N also allows us to estimate the density with the direct method.
In this case, we find an upper limit at $n_e \le 1~\rm cm^{-3}$ typical for the low density regime,
which is again consistent with our spatially-resolved measurement. 

In the nucleus, the S/N of both [SII] lines is high (see Figure \ref{reg:spect}) and the density is in the 
intermediate regime where the doublet line ratio becomes a useful diagnostic of the $n_e$. We can
therefore use the direct method, obtaining $n_e \sim 10^2~\rm cm^{-3}$.

With an estimate for the density, we then compute the recombination time-scale for the ionized gas:
\begin{equation}
{\tau_r  = \frac{1}{n_e \alpha_A} \sim \frac{10^5}{n_e} {\rm yr} }
\end{equation}
where $\alpha_A$ is the total recombination coefficient. At the high densities of the nucleus, 
the gas would recombine in less than $10^3$ yr, and the observed emission arises from on-going star formation 
that keeps the gas ionized. On the other hand, the gas in the tail can remain ionized up to 
$\sim 1$ Myr following ionisation. It is therefore interesting to consider whether the gas is being 
stripped as ionized and recombines in the tail, or whether other ionisation mechanisms are needed 
to power the observed emission. The orbital study of \citet{jac14} suggests an infall velocity of $~3000 \rm~km~s^{-1}$, 
which means that material at $\sim30$ kpc has been stripped at least 10 Myr ago, assuming 
instantaneous acceleration of the gas. Thus, according to our crude estimate, only the inner 
part of the tail could be recombining after being stripped in a ionized phase.
Other excitation mechanisms are at work in the full extent of the tail,
as shown by simulations \citep{tonnesen10}. Clearly, the uncertainties on the density and thus in the 
recombination timescale hamper firm conclusions. Arguably, both recombination of stripped gas 
in the ionized phase and from gas excited in situ within the tail are powering the observed emission.

\subsubsection{Electron temperature} \label{sectempden}

The high S/N of the composite spectra allows the detection of emission lines that are suitable to 
measure the electron temperature of the gas in the galaxy and its tail.
In principle, a few transitions like [OIII], [NII], and [SIII] could be used for this purpose, 
as emission lines from levels with different excitation energies fall into the optical window. 
However, given the wavelength coverage of our MUSE data, we are restricted to the 
use of [NII] for this calculation. 

We use equation (5.5) in \citet{ost06} and we assume the density dependence in the 
denominator is negligible. The error introduced in making this assumption is $0.4\%$ 
in the nucleus where $n_e \sim 10^2~\rm cm^{-3}$, and smaller elsewhere. The gas
temperature is given by
\begin{equation}
{T_e  = \frac{2.5 \times 10^4~{\rm K}}{\ln(0.121\times R_{\rm [NII]})}}
\end{equation}
where $R_{\rm [NII]} = \rm{[NII]}\lambda6584,6548/\rm{[NII]}\lambda5755$. 
In this calculation we use observed line ratios
not corrected for dust extinction, primarily because of our inability to derive the 
extinction curve especially in the tail and blob, where the ionisation mechanisms are
more uncertain. The values of $T_e$
must therefore be taken as lower limits, given that $R_{\rm [NII]}$ may in fact be 
lower then the value we measure. However, we checked that in the disc of the galaxy, where there 
is substantial extinction ($A_V \sim 1.44$ mag) the temperature would increase by only $\sim 6\%$. 

Clearly, the accuracy of this direct method is limited by the detection of the faint 
[NII]$\lambda5755$ line. In our composite spectra, we obtain marginal detections in all the regions, 
except the front. Fortunately, the wavelength where this line falls is clean from skylines 
and we can attempt a measurement of the temperature even in a moderate S/N regime. 
Robust fits are obtained by fixing the line width and centroid for all the [NII] lines, therefore
avoiding spurious solutions for the faint [NII]$\lambda5755$ line. Bootstrap errors are used to 
derive the uncertainties for the electron temperatures.
$T_e$ are then $(0.96 \pm 0.19) \times 10^4~$K in the galactic disk, 
$(1.8 \pm 0.3) \times 10^4~$K in the stacked HII regions and $(2.1 \pm 0.9) \times 10^4~$K, 
$(2.0 \pm 0.5) \times 10^4~$K for the tail and the blob respectively. 
At face value, this measurement suggests that the gas in \eso\ has temperatures commonly seen in 
photoionized regions, perhaps with a hint of a lower value in the galactic disc
compared to the tail or the HII regions. However, given the large uncertainties on the line flux 
and to lesser extent on the wavelength-dependent dust extinction, these values must be taken 
with caution.
Deeper observations are thus required to confirm this result with higher S/N detection 
of the weak [NII]$\lambda5755$. 


\subsubsection{BPT diagnostics}\label{sect:BPTdiag}

\begin{figure*} 
\includegraphics[scale=0.90]{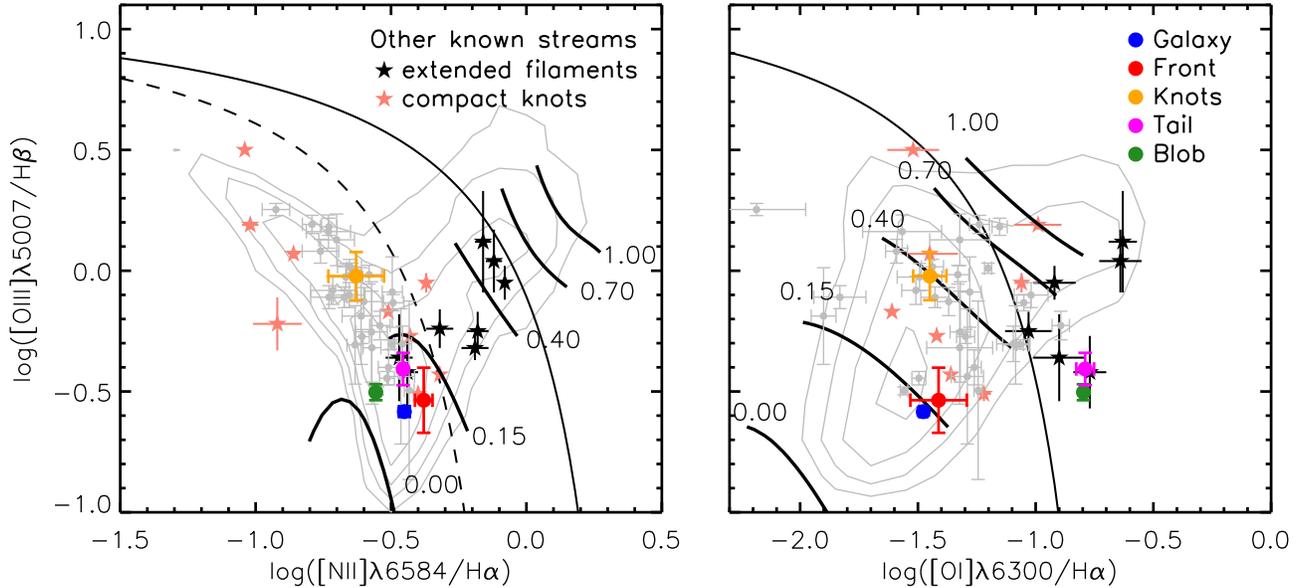}
\caption{BPT diagrams for the composite spectra extracted from the 5 characteristic regions of \eso.
Data points are colour coded as in Figure \ref{reg:spect}. Individual knots with robust detections 
in all the line ratios are also shown with small grey circles. The error bars shown for the composite 
knot spectrum reflect the sample variance, while the errors on the mean line ratios are of the size of the datapoint.
Black (pink) stars are extended filaments (compact knots) from \citet{yagi07}, \citet{yoshida12}, and \citet{merluzzi13}.
The grey contours are obtained from the nuclear spectra of a random sample of Sloan Digital Sky Survey (SDSS) 
galaxies at $0.03<z<0.1$.  
The solid black line is the extreme starburst separation between photoionisation and AGN/shock heating from \citet{kewley01}.
The dashed black line is the empirical division from \citet{kauffmann03} based on SDSS data. Thick solid lines
highlight different fractions (from 0 to 1) of \ha \ flux contributed by shocks from models by \citet{rich11}.
}\label{BPTdiag}
\end{figure*}

In section \ref{lineratiomaps}, we have shown two dimensional maps of various line ratios. We now use 
the higher S/N composite spectra to study emission line ratios from the detections of
 \hb, [OIII], [OI], [NII] and \ha\ in the 5 characteristic regions of \eso, defined in 
Figure \ref{reg:img}. Figure \ref{BPTdiag} shows the BPT diagrams for these composite spectra, 
using the same colour coding as in Figure \ref{reg:spect}. The 30 individual HII regions with detections in 
all the line ratios are also shown as small grey points. The error bars for the HII regions reflect
the sample variance, while errors on the mean line ratios are of the size of the datapoint.
The grey contours are obtained from the nuclear spectra of a random sample of SDSS galaxies
selected from the DR12 \citep{alam15} database at $0.03<z<0.1$, while 
the solid black line is the extreme starburst separation between photoionisation and 
AGN/shock heating from \citet{kewley01}. The dashed black line is the empirical division from 
\citet{kauffmann03} based on SDSS data. The thick solid lines
highlight different fractions (from 0 to 1) of \ha \ flux contributed by shocks from models by \citet{rich11}.
Moreover, in Figure \ref{BPTdiag} we show the line ratios obtained from the literature for extended
ionized gas filaments (black stars) and compact star forming knots (pink stars). \citet{yagi07}
observed the tail behind D100, a galaxy in the Coma cluster. We average the line ratios of their 
slits along the tail, weighting by the \ha \ flux. \citet{yoshida12} observed several knots and extended 
diffuse emission filaments behind two galaxies infalling into the Coma cluster (IC4040 and RB199).
Lastly, \citet{merluzzi13} observed a ram pressure stripped galaxy in the Shapley Supercluster, and 
we show the line ratios of their extraplanar ionized gas regions.

From the left panel, it is clear that all the HII regions in the tail as well as the central 
galaxy reside in the locus occupied by gas photoionized by stars. These HII regions span a broad range of
values in [OIII]$\lambda 5007/$\hb, which is indicative of different ionisation parameters.  This result is 
consistent with the line ratios of HII regions observed in the tails of other ram pressure stripped galaxies from the
literature (pink stars in Figure \ref{BPTdiag}).
Thus, despite the extreme ram pressure stripping suffered by \eso, the star formation properties of the galaxy 
disc and of the HII regions embedded in the tail appear typical. Indeed, the stacked spectrum of the 
galactic disc is well within the region where most of local galaxy nuclei are found. 
It should be noted, however, that this analysis is restricted to the most bound parts of the \eso\ disc, 
as the star formation activity is completely truncated in the outer parts of the disc, where ram pressure has already 
removed the galaxy ISM and no \ha \ emission is found (see paper I).

The galaxy front stands out as the region with the strongest [NII] contribution, which is however weak 
compared to the emission expected for a LINER or gas that is purely shock heated. This evidence, combined 
with the weak [OI] complicates the interpretation of which excitation mechanism is responsible for the gas
emission. Indeed at most 15\% of the excitation can be contributed by shocks in the front, 
according to the models by \citet{rich11}.
Moreover, the gas is kinematically cold in this region \citep[see][]{pap1}, 
perhaps surprising in presence of strong shocks moving through the gas. 
We therefore conclude that this region is mainly ionized, with an elevated 
[NII]/\ha \ ratio but a faint [OIII] emission indicative of a weak ionisation parameter.
As discussed above, however, we cannot rule out the contribution of shocks, as 
the separation between a putative thin shock-dominated region and the bulk of the photoionized ISM in the
\eso\ disc is poorly resolved in our observations. Moreover, composite spectra from 
extended regions are light weighted, and thus the measured fluxes near density 
discontinuities tend to be biased towards the brightest regions.

On the other hand, the tail and the bright blob in the outer tail show a contribution from 
very strong [OI] emitting gas, much beyond the locus typical for local galaxies. 
This evidence is consistent with the strength of [OI] in all the other ram pressure stripped tails
observations collected form the literature. However, the [NII]/\ha\  ratio is weaker compared to 
most of the other tails, a fact that is difficult to reconcile with 
ordinary shock-heating models. We note that the tail of
\eso \ hosts several HII regions, especially in the secondary tail where lower values of [NII]/\ha\ can be found. 
In this case, we are tempted to conclude that such peculiar line ratios in the tail arise 
from a combination of recombination from photoionized gas, (including gas that was stripped as ionized from the 
galaxy or that is ionized in situ by the HII regions), plus a shock contribution from the turbulence 
of the gas itself. This scenario is further supported by the high velocity dispersion of the 
gas along the tail \citep{pap1, jac14}.

 \begin{table*}
\caption{Fluxes measured in the composite spectra of the five regions shown in Figure \ref{reg:spect}. 
The values are normalised to the H$\beta$ flux and are not corrected for dust extinction. We also report the
dust extinction in units of $A_V$, measured from the Balmer decrement for the regions dominated by photionisation. }\label{tab:ratios} 


\centering

\begin{tabular}{l c c c c c c c }
\hline
\hline
     Line & $\lambda_{\rm r}$     &  $\Delta_v^{a}$    &    Region 1 &    Region 2  &    Region 3  &    Region 4 &    Region 5  \\
             &  (\AA)      & $\rm (km~s^{-1})$  &   (Disc) &     (Front)   &  (Knots) &   (Tail)   &  (Blob)   \\
\hline  
  H$\beta$       & 4861.33 &    0            & 1.000$\pm$0.022  & 1.000$\pm$0.065  & 1.000$\pm$0.013   & 1.000$\pm$0.064  & 1.000$\pm$0.020      \\ %
  \ion{[O}{III}]   &  4958.91 &  $<$9 (3)    & 0.095$\pm$0.006  & $<$0.189               & 0.322$\pm$0.066   & 0.111$\pm$0.058  & $<$0.141           	 \\ %
  \ion{[O}{III}]   &  5006.84 &  $<$10 (4)  & 0.261$\pm$0.007  & 0.252$\pm$0.102  & 0.952$\pm$0.174   & 0.392$\pm$0.060  & 0.314$\pm$0.023       \\ %
  \ion{[N}{II}]    &  5754.64 & $<$58 (3)  & 0.016$\pm$0.009  & $<$0.135               & 0.034$\pm$0.008   & 0.065$\pm$0.045  & 0.047$\pm$0.019        \\ %
  \ion{[O}{I}]     &  6300.30 & $<$24 (3)  & 0.151$\pm$0.005  & $<$0.173               & 0.145$\pm$0.010   & 0.694$\pm$0.060  & 0.636$\pm$0.018     \\ %
  \ion{[O}{I}]     &  6363.78 & $<$19 (1)  & 0.050$\pm$0.005  & $<$0.147               &    $<$0.051             & $<$0.138               & 0.269$\pm$0.026     \\ %
  \ion{[N}{II}]    &  6548.05 & $<$22 (3)  & 0.501$\pm$0.017  & 0.322$\pm$0.025  & 0.287$\pm$0.056   & 0.467$\pm$0.024  & 0.357$\pm$0.010     \\ %
  H$\alpha$     & 6562.82  & $<$20 (1)  & 4.529$\pm$0.092  & 2.998$\pm$0.038  & 3.954$\pm$0.140   & 4.267$\pm$0.040  & 3.984$\pm$0.019    \\ %
  \ion{[N}{II}]    &  6583.45 & $<$24 (3)  & 1.604$\pm$0.065  & 1.190$\pm$0.036  & 0.928$\pm$0.106   & 1.499$\pm$0.032  & 1.107$\pm$0.013     \\ %
  \ion{[S}{II}]    & 6716.44  & $<$23 (5)  & 0.921$\pm$0.022  & $<$0.174               & 0.745$\pm$0.039   & 1.629$\pm$0.116   & 1.056$\pm$0.015     \\ %
  \ion{[S}{II}]    & 6730.81  & $<$20 (5)  & 0.651$\pm$0.016  & $<$0.219               & 0.470$\pm$0.037   & $<$0.297               & 0.422$\pm$0.030    \\ %
  \ion{[S}{III}]   & 9068.60  & $<$16 (2)  & 0.249$\pm$0.010  & 0.210$\pm$0.055  & 0.241$\pm$0.030   & $<$0.216               &    $<$0.108	 	    \\ %
\hline   
  $A_V$            &               & & 1.44$\pm$0.09  & - & 1.02$\pm$0.25   & -  & -   \\ %
\hline
\end{tabular}

\flushleft{$^a$ Absolute value of the largest offset peak centre relative to H$\beta$. The value in parenthesis indicates the region where this occurs.}
\end{table*}

\subsection{Properties of the HII regions}\label{sec:HII}

In this section we estimate the density, metallicity and ionisation parameters  
for regions where the gas is mainly photoionized by young stars, namely the HII regions 
embedded in the tail and the body of the \eso\ galaxy. 

\begin{figure*} 
\includegraphics[scale=0.90]{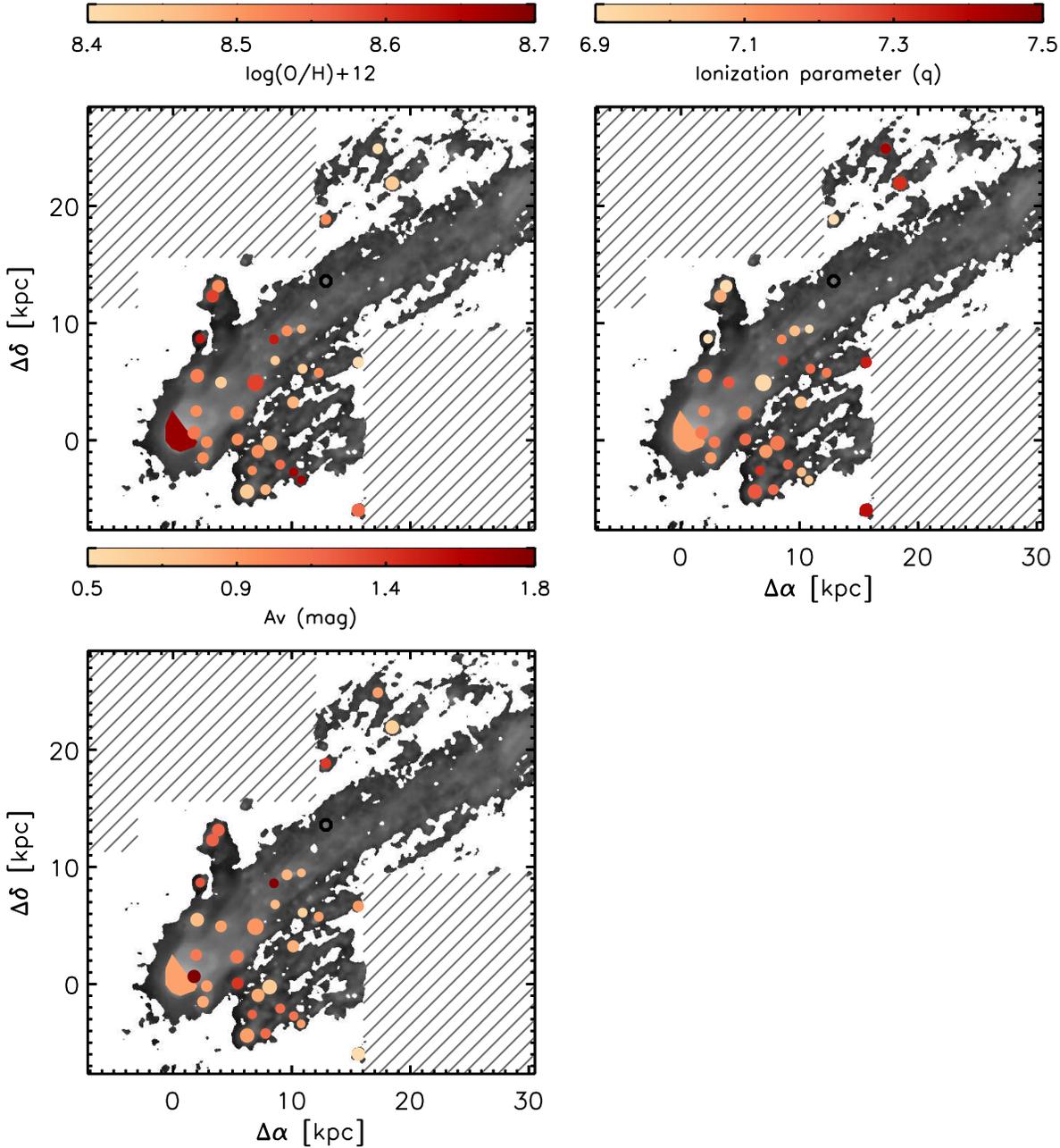}
\caption{Top: maps of the gas phase metallicity and ionisation parameter for the HII regions and the disc of 
\eso\ obtained from nebular emission lines fitting with the {\sc izi} code. Bottom: map of the dust extinction in the 
same regions, computed from the Balmer decrement. One HII region with insufficient S/N in \hb\ is shown as an 
empty black circle. The \ha \ diffuse emission is shown in grey scale as a reference for the position of the HII regions
along the tail. } 
\label{fig:physicsHII}
\end{figure*}

As a first step, we correct the emission line spectra for dust extinction using the Balmer decrement. 
Having restricted our analysis to purely photoionized regions, and with a temperature of $T\sim  10^4~$K, 
we can assume a theoretical \ha/\hb\ ratio
from case B recombination. This value depends only weakly on density and temperature and is 
2.86 in the low density limit at $T=10^4K$ \citep{ost89}. 
In case of density and temperature variations, the variation in the theoretical \ha/\hb \ ratio is minimal.
Therefore the variations in the observed ratios can be mainly attributed to variations in the dust content.
We compute the Balmer decrement as:
\begin{equation}
{C(H\beta) = \frac{\log(2.86) - \log[\frac{F(H\alpha)}{F(H\beta)}]_{\rm obs}}{f(H\alpha)}}
\end{equation}
where $f({\rm H\alpha})=-0.297$ is the selective extinction of \ha\ relative to \hb\ from the Galactic extinction 
law of \citet{cardelli89}, which we also use for the correction of all the observed emission lines.
Under these assumptions, we express the attenuation in units of $A_{V}= 2.5 \times 0.86 \times C({\rm H\beta})$.
A map of the extinction of the HII regions is given in the bottom panel of Figure \ref{fig:physicsHII}. 
The extinction is on average 1~mag, with substantial scatter from region to region, suggestive of 
varying dust content within the body and tail of this galaxy (and perhaps along the line of sight to it).
No clear trends are visible with the position along the tail.

The density for the HII regions in the tail, can be estimated following eq. \ref{eq:Lha} but adopting a spherical 
geometry in which the volume is $V=4\pi R^3/3$. In our observations, we cannot resolve the size of the individual 
HII regions and we assume a constant radius  $R=20~\rm pc$ \citep[see][]{gutierrez10}.
Also for the HII regions, the filling factor is unknown and likely to be a function of radius \citep{cedres13}.
In this case, we assume $f=1$ meaning that our estimates provide a lower limit for the local density of 
these HII regions. Although the uncertainty on the density is very large, we find $n_e \gtrsim 10^2~\rm cm^{-3}$, 
at least two orders of magnitude denser than the diffuse gas in the tail. 

With the dust corrected spectra, we can infer the chemical abundance of the HII regions 
using photoionisation modelling. The metallicity is of great importance to understand
how the star formation proceeds in stripped tails, and how much of the stripped gas pollutes the 
intracluster environment. Another interesting parameter, a by-product of the metallicity analysis, 
is the ionisation parameter ($q$), defined as the flux of ionising photons per unit gas density.

In order to derive the gas phase metallicity  ($O/H$) and ionisation parameter of the gas, 
we use the {\sc izi} \citep{blanc15} code to fit the observed nebular emission lines in our spectra against
photoionisation models. For this analysis, we use the input models from \citet{kewley01},  
who derive line fluxes from the {\sc mappings-III} code with input stellar spectra computed using the 
{\sc starburst99} code \citep{lei99}. 
Those models are computed for a gas density $n_e = 350~\rm cm^{-3}$, a constant star formation rate,
and they assume a Salpeter \citep{salpeter55} initial mass function (IMF). The {\sc izi} code uses Bayesian statistics 
to derive the joint marginalised probability density function (PDF) for $O/H$ and $q$. 
Here, we run the code with flat priors, with the exception of a handful of problematic cases.
Indeed, during the analysis we identify 4 HII regions for which the posterior PDF has multiple peaks for
at least one of the parameters of interest. In these few cases, we impose a Gaussian prior on $q$ 
which is motivated by the excellent linear correlation ($r=0.91$) between  $q$ and the [OIII]$\lambda5007$/\hb\ 
ratio observed in the remaining HII regions where the PDFs are well behaved. 
We checked that the results do not change if we assume models with density $n_e = 10~\rm cm^{-3}$,
as data constrain only the ionisation parameter with a degeneracy between density and 
intensity of the ionising radiation.

Maps of $O/H$ and $q$ are shown in the top panels of Figure \ref{fig:physicsHII}. 
We start by noting that the metallicity in the body of the galaxy is $\log(O/H)+12\sim8.65$, 
which is consistent with the determination $\log(O/H)+12\sim8.7$ by \citet{sun07}, 
obtained using different data but the same suite of models. This agreement confirms 
the lack of systematic biases due to flux calibrations in the analysis. 
The uncertainty on the metallicity of individual HII regions is $~0.1$dex as obtained from the
widths of the posterior distributions. However, the use 
a specific grid of models introduces an additional systematic uncertainty, 
which we estimate around $~0.15$dex error by repeating the analysis with different models. 
Therefore the metallicity is known with a quite large uncertainty. 

The spatial variations we find in Figure \ref{fig:physicsHII} are small and consistent with constant metallicity 
throughout the tail with an average value of $\log(O/H)+12\sim8.55\pm0.06$, which is 0.75 times the value assumed 
for the Solar metallicity. A similar behaviour is seen for the ionisation parameter, although 
uncertainties in individual measurements hamper firm conclusions on a spatial variation, 
if any is present.

Lastly, we convert the \ha\ de-reddened flux into a SFR using the \citet{kennicutt98} calibration. 
The total SFR of the selected HII regions is $SFR = 0.62\pm0.03 \rm{M_\odot~yr^{-1}}$, 
in excellent agreement with the value $SFR = 0.59 \rm{M_\odot~yr^{-1}}$ found by \citet{sun07} using 
the same SFR calibration and a Salpeter IMF. This confirms once again the quality of the absolute flux calibration of the 
MUSE observations. We caution however that, because of the low intrinsic SFR of individual HII regions,
a full treatment of the stochastic IMF sampling is likely required for a more precise measurement of the 
SFR \citep{das14}.

While there is no systematic trend of the physical properties of these HII regions with respect to their 
spatial location, the distribution of HII regions is in itself peculiar and traces the imprint of
the formation mechanism. One third of the HII regions are found in leading edge of the secondary tail
and they are among the brightest that we have observed in \eso. Other HII regions are distributed 
along the main tail and only a few of them are found at a distance greater than 15 kpc from the galactic disk. 
Moreover, most of the HII regions are located in the inner part of the tail, where the velocity dispersion 
of the gas on scales larger than individual HII regions is low ($\sigma \sim 25-50$ \kms). In fact, the most 
distant HII regions are not found in the primary tail, 
where we observe an increased velocity dispersion (paper I). Rather, they reside in the diffuse gas in the 
northern part of our field of view, where the velocity dispersion is indeed low. If HII regions were formed 
in the galactic disc and stripped downstream by ram pressure, then they would be expected to lie 
anywhere in the tail. However, if instead the HII regions are formed in situ within the tail, then 
they are preferentially expected to reside where the gas is less turbulent, and the gas can cool
and self-shield to form molecules. 

This argument is complicated by age effects, as the stripped HII regions would progressively 
fade as the gas is accelerated to larger distances from the galaxy body. Further, some of the HII regions
may not survive the violent stripping and hydrodynamic interactions with the hot ICM. 
However, another piece of evidence in favour of in situ formation, at least for some HII regions,
comes from the analysis of \citet{sun07}, who reported young ages $<8$ Myr for all the HII regions
considered. If they were formed in the galaxy, then a clearer age gradient would be expected, 
as the galaxy moves in the cluster potential and the HII regions trail behind it. As we noted, 
also \citet{sun07} caution that older HII regions may have faded in the outer tail, below 
the depth of our observations.
 
\section{Summary and Conclusions}

In this paper, we have presented the analysis of the physical properties of the disc and stripped tail of \eso, 
a galaxy that is infalling into the nearby Norma cluster at high velocity. Leveraging on the sensitivity and 
spatial resolution of MUSE observations, we have studied for the first time 2D maps of emission line ratios 
within a posterchild of violent ram-pressure stripping. 

Due to the complex interaction between different gas phases, several physical processes are likely involved in 
the ionisation that powers the observed emission of the tail and the embedded HII regions.
While we cannot fully disentangle each source of ionisation, our analysis has provided 
new insight into the physical conditions of the emitting gas. Our results can be summarised as follows.

\begin{itemize}
\item[--]  Considering the tail, the analysis of line ratios shows high values of [NII]/\ha\ and [OI]/\ha.  
Thus, photoionisation alone cannot be responsible for the observed emission, and the contributions of 
other mechanisms is required. A contribution from shocks is likely, also given the turbulent nature 
of the gas in the outer tail, but we cannot exclude thermal conduction with the hot ICM and 
MHD waves as other excitation mechanisms.
At the same time, line ratios are not as strong as commonly seen in pure shock heating models, and the gas temperature
of $(2.1 \pm 0.9)\times 10^4~$K is similar to what found in photoionized gas. 
Part of the emission, particularly at close separations from the galaxy, may originate from 
recombination of photoionized gas stripped from the main body of the galaxy. The electron density in the 
tail is indeed low, $n_e \sim 0.3~\rm cm^{-3}$, and thus the recombination timescale is long enough to 
keep the gas ionized up to a few kpc downstream.

\item[--] The bulk of the front region where the galaxy ISM first collides with the hot ICM is 
not dominated by strong shocks, as [OI] is only weakly or mostly undetected in our observations. 
Line ratios are mostly consistent with photoionized gas, although with a higher [NII]/\ha\
ratio compared to the main body of the galaxy. We caution, however, that the front region is poorly 
resolved in our observations, and the separation with the bulk of the galaxy ISM is ambiguous.

\item[--] A large number of bright compact knots have been identified in the tail. These regions stand out 
because of their high [OIII]/\hb\ and low [NII]/\ha\ and [OI]/\ha\ ratios, fully consistent with the locus 
occupied by HII regions in the BPT diagram. These HII regions have 
densities $n_e \gtrsim 10^2~\rm cm^{-3}$ and temperatures $(1.8 \pm 0.3)\times 10^4~$K.
Thus, despite residing in ram-pressure stripped tails, they exhibit usual properties, commonly observed 
in HII regions. By comparing the line ratios to a grid of photoionisation models, 
we found a metallicity close to the solar value ($\sim 0.7$ solar), albeit with substantial scatter.

\item[--] By comparing the spatial position of the HII regions to the kinematic maps presented in paper I, we 
found the HII regions are preferentially located in regions where
the gas is dynamically cool, with a velocity dispersion less than 25 - 50 \kms. Indeed no HII regions are found 
in the outer part of the tail where \citet{pap1} and \citet{jac14} found a high degree of turbulence. 
Moreover, the young ages of these regions as well as the lack of any trend in the physical properties 
as the function of their position suggest that these HII regions have formed in situ 
from the stripped gas in the tail, as also concluded by previous studies of this and 
similar systems \citep{sun07, arrigoni12, fossati12}. However, we can not fully 
exclude selection biases that prevent us from detecting older HII regions stripped from the main body. 

\end{itemize}

Our analysis represents the first attempt to characterise the rich physics of the 
\eso\ tail. In this work, we have primarily focused on the most common emission line diagnostics 
for shocks and photoionized gas. However, other processes such as thermal conduction and MHD waves
may play a role in the excitation of the tail.
Deeper observations of the full extent of the tail that have been recently 
obtained with MUSE, together with detailed radiative transfer models, will 
allow future explorations of other diagnostics (possibly including weak lines such as [Fe X]) 
that are necessary to gain a more complete understanding of the physics at work in 
ram-pressure stripped tails.  

Our study demonstrates the feasibility of detailed analyses of the properties of diffuse low-surface 
brightness material with large-field-of-view IFUs on 8-10 m class telescopes. By targeting 
galaxies like \eso, future observations will be able to provide key information on how gas is 
ablated from galaxies that are infalling into clusters of different richness and along different orbits.
Moreover, by examining spatially resolved line ratios, these studies will inform on what 
mechanisms power the observed emission. In turn, this will shed new insight in the 
physics of star formation within the stripped material, a possible source of intracluster light and 
metals for the ICM.  Future observations that target cluster galaxies or groups across a wide range of
environments will thus yield a comprehensive view of how ram pressure shapes the formation of the red 
sequence in the local Universe.

\section*{Acknowledgments}
It is a pleasure to thank Giulia Mantovani, Eva Wuyts, Trevor Mendel, Mark Krumholz, Audrey Galametz and 
Gary Ferland for useful discussion. We thank the anonymous referee for his/her constructive comments.
This work is based on observations made with ESO telescopes at 
the La Silla Paranal Observatory under programmes ID 60.A-9349(A) and ID 60.A-9303(A). 
M. Fossati and DJW acknowledge the support of the Deutsche Forschungsgemeinschaft 
via Project ID 3871/1-1. M. Fumagalli acknowledges support by the Science and Technology 
Facilities Council [grant number  ST/L00075X/1].
For access to the data used in this paper, the {\sc kubeviz} line fitting software, and the IDL codes to 
process the MUSE data cubes, please contact the authors.

\label{lastpage}

\end{document}